# Two-photon IR Pumped UV-Vis Transient Absorption Spectroscopy of Dirac Fermions in the 2D and 3D Topological Insulator $Bi_2Se_3$

**Yuri D. Glinka** [1,2,*], **Tingchao He** [3,*] **and Xiao Wei Sun** [1,4,*]

[1] Guangdong University Key Lab for Advanced Quantum Dot Displays and Lighting, Shenzhen Key Laboratory for Advanced Quantum Dot Displays and Lighting, Department of Electrical and Electronic Engineering, Southern University of Science and Technology, Shenzhen 518055, China
[2] Institute of Physics, National Academy of Sciences of Ukraine, Kyiv 03028, Ukraine
[3] College of Physics and Energy, Shenzhen University, Shenzhen 518060, China
[4] Shenzhen Planck Innovation Technologies Pte Ltd., Longgang, Shenzhen 518112, China

[*] Correspondence: (YDG) yuridglinka@yahoo.com, (TH) tche@szu.edu.cn, (XWS) sunxw@sustech.edu.cn

**Abstract:** It is often taken for granted that in pump-probe experiments on the topological insulator (TI) $Bi_2Se_3$ using IR pumping with a commercial Ti:Sapphire laser [~800 nm (1.55 eV photon energy)], the electrons are excited in the one-photon absorption regime, even when pumped with absorbed fluences in the mJ cm$^{-2}$ range. Here, using UV-Vis transient absorption (TA) spectroscopy, we show that even at low-power IR pumping with absorbed fluences in the μJ cm$^{-2}$ range, the TA spectra of the TI $Bi_2Se_3$ extend across a part of the UV and the entire visible region. This observation suggests unambiguously that the two-photon pumping regime accompanies the usual one-photon pumping regime even at low laser powers applied. We attribute the high efficiency of two-photon pumping to the giant nonlinearity of Dirac fermions in the Dirac surface states (SS). On the contrary, one-photon pumping is associated with the excitation of bound valence electrons in the bulk into the conduction band. Two mechanisms of absorption bleaching were also revealed since they manifest themselves in different spectral regions of probing. These two mechanisms were assigned to the filling of the phase-space in the Dirac SS and bulk states and the corresponding Pauli blocking.

## 1. Introduction

Ever since 3D topological insulators (TIs) were first discovered [1-3], the energy relaxation dynamics of photoexcited carriers in these materials has been extensively studied during the last decade or so using a variety of pump-probe techniques exploiting ultrashort IR pulses from a commercial Ti:Sapphire laser [~800 nm (1.55 eV photon energy)] [4-25]. The increased interest in ultrafast carrier dynamics is completely governed by the photonic/plasmonic properties of TIs, which are promising for optoelectronics and spintronics in the THz range [26-31]. Consequently, time- and angle-resolved photoemission spectroscopy (TrARPES) [4-12], pump-probe reflectivity [13-18], pump-probe second harmonic generation (SHG) [14,19], pump-probe X-ray diffraction [20], transient reflectivity spectroscopy [21-23], and transient conductivity spectroscopy [24,25] were used for these purposes. It is important to note that, compared to conventional one-photon ARPES, TrARPES uses exclusively two-photon photoemission. Specifically, the first photon (usually in the mid-IR, IR, visible or even UV range) creates a distribution of pump-excited electrons in the initially unoccupied Dirac surface states (SS) and bulk states, and the second photon (usually in the UV range) maps this distribution with momentum resolution through angle-resolved photoemission [4-12]. The sum of the energies of two photons must at least exceed the work function of the sample, 5.97 eV [12]. However, it is generally assumed that pumping in TrARPES experiments occurs in the one-photon absorption regime, as in all pump-probe experiments using reflectivity, SHG, X-ray, and conductivity as probing methods. Moreover, this statement is considered true even if the carriers were pumped at high powers with absorbed fluences in the mJ cm$^{-2}$ range [14,19,20,25].

On the contrary, recent applications of transient absorption (TA) spectroscopy to study ultrathin $Bi_2Se_3$ films have clearly demonstrated that for the IR pumping regime, the two-photon excitation of massless Dirac fermions in Dirac SS must also be considered on a par with one-photon pumping of the bound valence electrons in the bulk states [32]. Moreover, the relaxation dynamics observed for the Dirac SS using two-photon IR pumping turned out to be very close to that observed with the corresponding one-photon UV pumping [33,34], thus unambiguously confirming the reality of two-photon excitation. Like TrARPES, TA spectroscopy also monitors the distribution of electrons excited by pumping photons but using the Burstein-Moss effect [35–37] applied to all allowed electronic states [38]. In this case, the momenta of the electrons remain unresolved. Consequently, one-photon pumped TA spectroscopy can only be realized at energies of pumping and probing photons below the work function of the sample to avoid photoemission. This behavior also means that the pumping photon energy for two-photon pumped TA spectroscopy of the TI $Bi_2Se_3$ should be lower than ~2.98 eV, and the probing photon energy should not exceed ~5.97 eV. Despite this fundamental difference between TrARPES and two-photon pumped TA spectroscopy, both seem to complement each other.

In this article, we present a thorough study of relaxation dynamics of surface Dirac fermions and bulk electrons, including their mass gain/loss dynamics upon relaxation, in the 2D and 3D TI $Bi_2Se_3$ using TA spectroscopy exploiting low-power IR pumping with absorbed fluences in the µJ cm$^{-2}$ range. All measurements were carried out using ultrashort (~100 fs) pumping pulses at a wavelength of 730 nm (photon energy 1.7 eV) and probing in the UV-Vis region (1.65 - 3.8 eV). The corresponding film thicknesses were 2 and 10 quintuple layers (QL) stacked together by van der Waals interaction (~2 and ~10 nm thick films, respectively), where QL represents five covalently bonded Se-Bi-Se-Bi-Se atomic sheets. Although, according to ARPES, the transition between 3D and 2D TI $Bi_2Se_3$ occurs at 6 QL [39] and TA spectroscopy also confirms this behavior [32], we used a topologically trivial insulator phase (2 nm thick film) as the 2D TI $Bi_2Se_3$. This assignment seems reasonable, since the TA spectra of topologically trivial insulator phase and those of the gapped topologically nontrivial insulator phase (4 and 5 nm thick films) turned out to be qualitatively similar [32,34]. We have shown that even at low pumping power, TA spectra extend across a part of the UV and the entire visible region. This behavior suggests unambiguously the presence of two-photon pumping. Consequently, we assumed that TA spectra directly image the relaxation dynamics of two-photon-pumped Dirac fermions, which occurs simultaneously with the usual relaxation of one-photon-pumped electrons in the bulk of the film. We associate the high efficiency of two-photon pumping with the giant nonlinearity of Dirac fermions in the Dirac SS. This conclusion is hence in good agreement with predictions for 2D Dirac systems, which are expected to exhibit an extremely strong non-linear optical response [40,41].

Analyzing the corresponding pump-probe traces as a function of probing photon energy, we also recognized two mechanisms of absorption bleaching (AB), which appear in different spectral regions. We associate them with the Burstein-Moss effect, which includes phase-space filling in the Dirac SS and bulk states and the corresponding Pauli blocking. We also have shown that since the TI $Bi_2Se_3$ is a van der Waals system, the coherent longitudinal optical (LO) phonon oscillations observed in the AB pump-probe traces of the 3D TI $Bi_2Se_3$ are transiently evoked by the vertical electron transport, which spatially synchronizes the electron-LO-phonon scattering events in individual QLs. As the vertical electron transport becomes negligible, as that occurs in the 2D TI $Bi_2Se_3$, the coherent LO-phonon oscillations in the AB pump-probe traces become unresolved, whereas they emerge in the pump-probe traces associated with the inverse bremsstrahlung type free carrier absorption (FCA) in the Dirac SS. The coherent LO-phonon oscillations in the FCA pump-probe traces are caused by the in-phase successive quasielastic scattering of Dirac fermions from the uppermost atomic layer of the surface. Finally, we concluded that the two-photon IR pumped UV-Vis TA spectroscopy of Dirac fermions is a powerful and selective tool for studying ultrafast relaxation in 2D Dirac systems.

## 2. Experimental details
### 2.1. Sample preparation
The 2 and 10 nm thick $Bi_2Se_3$ films were grown on 0.5 mm $Al_2O_3$(0001) substrates by molecular beam epitaxy, with a 10 nm thick $MgF_2$ protecting capping layer, which was grown at room temperature without exposing the film to the atmosphere. The samples have been found to be epitaxial and the nominal number of QL was accurate to approximately 5%. The level of disorder of the grown films and their quality were approximately the same [42]. Furthermore, the films reveal the thickness-dependent *n*-type doping with the free carrier density of ~$10^{19}$ cm$^{-3}$ [43].

### 2.2. Experimental setup.
TA spectra were measured using a TA spectrometer (Newport), which was equipped with a Spectra-Physics Solstice Ace regenerative amplifier (~100 fs pulses at 800 nm with 1.0 KHz repetition rate) to generate the supercontinuum probing beam and a Topas light convertor for the pumping beam. The spectrometer has also been modified to suppress all coherent artifacts emanating from the sapphire substrates and appearing on a subpicosecond time scale. We used optical pumping at 730 nm (1.7 eV photon energy) and a supercontinuum probing beam generated in a sapphire plate and spread across the UV-Vis spectral region from ~1.65 to ~3.8 eV. The probing beam was almost in normal incidence, while the pumping beam was at an incident angle of ~30°. All measurements were performed in air and at room temperature using a cross-linear-polarized geometry. Specifically, the pumping and probing beams were polarized out-of-plane (vertical) and in-plane (horizontal) of incidence, respectively.

The data matrix was corrected for the chirp of the supercontinuum probing pulse using a coherent nonlinear optical response from a thin (0.3 mm) sapphire plate (presumably degenerate four-wave mixing or two-photon absorption [44,45]). Although the amplitude of the coherent response was at least two orders of magnitude smaller than that of the actual pump-probe signal from the sample, the coherent peak is shifted over the entire probing region with delay-time and made it possible to observe the effect of supercontinuum pulse chirping [34], which is approximated using the parabolic equation [Fig. 1(f) and (g)] [45]. The corresponding zero-time for every wavelength was adjusted numerically in all datasets. The temporal pulse chirping effect was found to be much shorter than the actual subpicosecond relaxation dynamics of highly energetic two-photon-excited Dirac fermions, as discussed further below.

The spot sizes of the pumping and probing beams were ∼400 and ∼150 µm, respectively. The pumping beam average power was ∼0.5 mW, which corresponds to the pumping intensity (power density) of ∼4.0 GW cm$^{-2}$. The higher power pumping of ∼3.0 mW (∼27 GW cm$^{-2}$) was used for comparison with previously reported data [32]. We will refer these two pumping regimes further below to as the low-power and high-power pumping regimes. Because one-photon absorption in the bulk states and two-photon absorption in the Dirac SS occur simultaneously, the attenuation of the initial light intensity ($I_0$) propagating the distance $z$ through the Bi$_2$Se$_3$ film can be expressed as [38]

$$\frac{\partial I(z)}{\partial z} = -\alpha I_0(z) - \beta I_0^2(z), \tag{1}$$

where $a$ and $b$ are the linear (one-photon) and nonlinear (two-photon) absorption coefficients for the bulk states and Dirac SS, respectively. However, these two contributions seem to be undistinguished due to the extreme thinness of the Dirac SS. It is worth noting that owing to this two-component behavior, the effective linear absorption coefficient that was measured for thin Bi$_2$Se$_3$ films has revealed a ∼2-fold increase with decreasing film thickness from ∼40 to ∼6 nm [16]. Subsequently, this increase in the linear absorption coefficient has been found to be in good agreement with an increase in the free carrier density measured for the same samples using the Hall effect [43]. This correlation between an increase in the free carrier density and the linear absorption coefficient with decreasing film thickness unambiguously suggests that the two-photon absorption is associated with the free carrier population initially residing in the Dirac SS due to natural *n*-doping. As a result, we will use the effective linear absorption coefficient to estimate the corresponding absorbed fluences, instead of taking into consideration the simultaneously acting one-photon and two-photon absorption. Despite the imperfection of this procedure, it seems appropriate if we assume the comparable rates of one-photon and two-photon absorption.

The corresponding absorbed fluences for the low-power and high-power pumping regimes were estimated using the pumping light absorptivity (A = 1 – R – T), reflectivity (R), and transmissivity (T) at ∼0.21 mJ cm$^{-2}$ and ∼1.0 mJ cm$^{-2}$ for the 10 nm thick film and at ∼16 µJ cm$^{-2}$ and ∼27 µJ cm$^{-2}$ for the 2 nm thick film. Here we used the following measured parameters: the effective linear absorption coefficient of $\alpha$ = ∼1.0×10$^6$ cm$^{-1}$ and ∼1.2×10$^6$ cm$^{-1}$, as well as the reflectivity of R = ∼0.25 and ∼0.2, for the 10 and 2 nm thick films, respectively [16]. We note that the calculated fluences for the low-power pumping regime are slightly higher than those used in TrARPES and some pump-probe reflectivity experiments [4-12], whereas they are comparable or even lower those used in pump-probe measurements exploiting SHG, X-ray, and conductivity as probing methods [14,19,20,25]. The broadband probing beam was of ∼0.4 mW average power, which for the same as the pumping beam bandwidth (∼26 meV) provides the probing intensity of ∼0.15 GW cm$^{-2}$. Because the latter value is much smaller than that of the pumping beam, the probing beam effect on carrier excitation seems to be negligible.

3. **Results and discussions**
*3.1. The TA spectra of the 2 and 10 QL thick Bi$_2$Se$_3$ films.*

The TA spectra of the 2 and 10 nm thick Bi$_2$Se$_3$ films, representing the 2D and 3D TI Bi$_2$Se$_3$, respectively, which were measured using the low-power pumping regime, are extended from ∼1.65 to ∼3.8 eV [Fig. 1(a), (b) and (c), (d)]. This broadband contribution gradually develops on a subpicosecond time scale. The TA spectra of the 2D TI Bi$_2$Se$_3$ consists of negative and positive contributions, while only negative contributions appear for the 3D TI Bi$_2$Se$_3$. These spectral features are fully consistent with those previously reported for more powerful pumping [32]. Specifically, the positive contribution is known to be due to the inverse bremsstrahlung type FCA in the Dirac SS2, while the negative contributions are associated with the conduction band (CB) AB (the broadband negative contribution) and the valence band (VB) AB (the narrow negative contribution with a maximum almost at the pumping photon energy) [Fig. 1(e)]. Since the broadband CB-AB contribution significantly exceeds in energy the pumping photon energy, it clearly points to the two-photon pumping of Dirac fermions, initially residing in the upper cone of the Dirac SS1 below Fermi energy ($E_F$), toward the higher energy Dirac SS4 [Fig. 1(e)] [32-34]. The two-photon excitation transition is expected to occur resonantly due to direct optical coupling between the Dirac SS1 and the Dirac SS2 [12]. It should be noted that other bulk-to-surface and surface-to-bulk electronic interband transitions can be realized in the TI $Bi_2Se_3$ using the one-photon pumping regime with 3.02 eV photons [46]. It can be expected that similar interband transitions can also be realized with 1.7 eV photons. However, two-photon pumping seems to occur exclusively for surface-to-surface transitions due to the giant nonlinearity of massless Dirac fermions [40,41]. According to what we mentioned in the Introduction, two-photon pumping can be achieved only at photon energies not exceeding half the work function of the sample (2.98 eV). As a result, the TA spectroscopy of massless Dirac fermions at a pumping photon energy of 3.02 eV, used in Ref. 46, can only be implemented in the one-photon pumping regime. Furthermore, the corresponding two-photon pumping in TrARPES experiments will lead to photoemission even without applying a probing beam. Note also that the power dependence of the CB-AB response cannot be used to confirm the two-photon nature of the excitation, as, for example, in photoluminescence measurements [47], since the amplitude of the CB-AB response is not directly proportional to the density of photoexcited carriers [32,38].

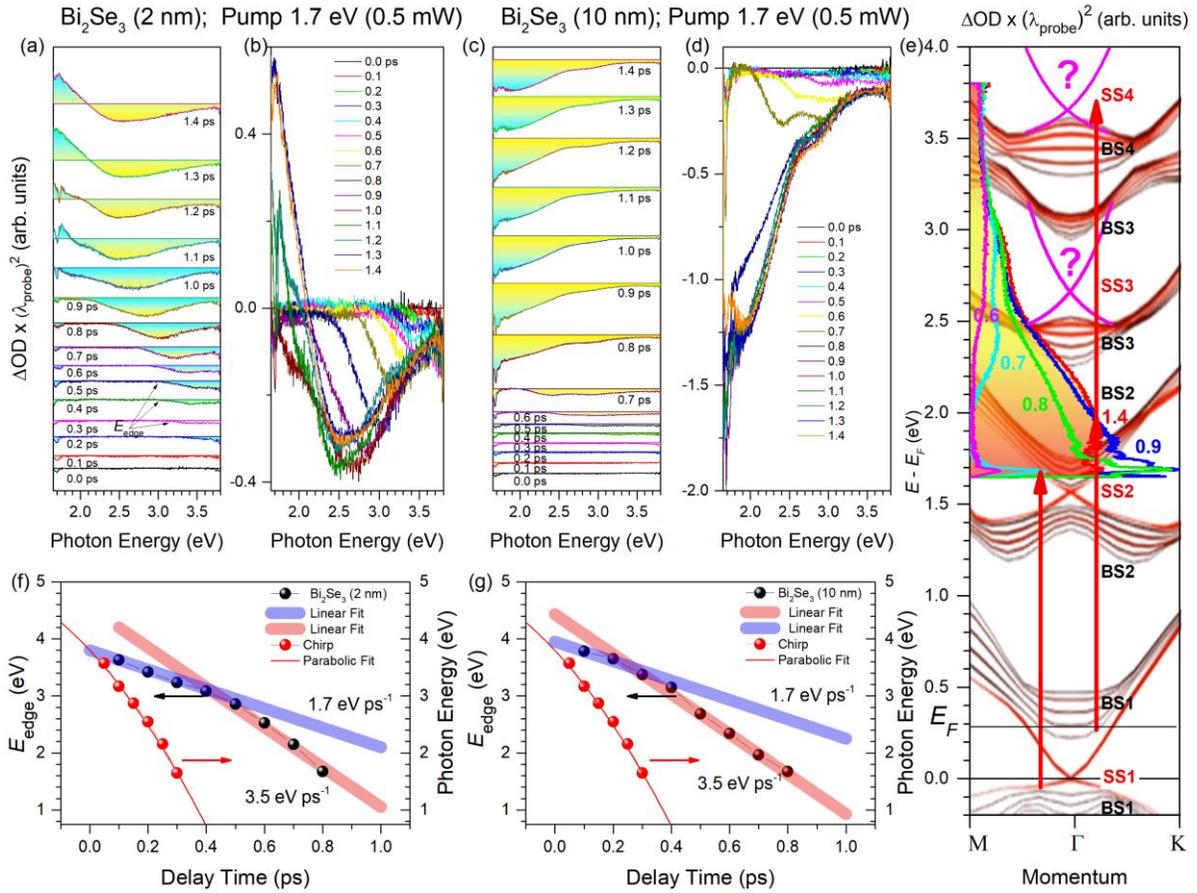

**Figure 1.** (a), (b) and (c), (d) Set of TA spectra of the 2D and 3D TI $Bi_2Se_3$ (2 QL and 10 QL thick films, respectively). The spectra were measured at delay times indicated by the corresponding colors using the ~730 nm pumping (~1.7 eV photon energy) of ~0.5 mW power (~4 GW cm$^{-2}$ pulse peak intensity, ~16 µJ cm$^{-2}$ and ~0.21 mJ cm$^{-2}$ absorbed fluence for 2 and 10 nm thick films, respectively). Parts (a) and (b), as well as parts (c) and (d), show the same TA spectra for clarity. The zero-intensity lines of TA spectra in parts (a) and (c) were shifted along the ΔOD axis (optical density change) for better observation. The factor $(\lambda_{probe})^2$ in the ΔOD axis arises due to wavelength-to-energy units' transformation. (e) Band structure of the 6 QL thick 3D TI $Bi_2Se_3$ film calculated in Ref. 12 and the one-photon and two-photon pumping transitions (red vertical up arrows) originating from the bulk valence band states and from Dirac surface states below Fermi energy ($E_F$), respectively. The bulk and Dirac surface states are marked as BS and SS, respectively. The predicted in Ref. 33 higher energy Dirac surface states are marked as "?". Some of TA spectra shown in part (c) and (d) are plotted on the band structure diagram, as indicated by the corresponding color numbers presenting delay-times in ps. (f) and (g) Temporal evolution of the low-energy edge of the transiently excited electron population [$E_{edge}$, as defined in part (a)] for the 2D and 3D TI $Bi_2Se_3$, respectively, and the chirp of the supercontinuum probing pulse. The results of the linear fit with the corresponding electron energy relaxation rates in units of eV ps$^{-1}$ and the parabolic fit of the chirp are shown.

Further relaxation of two-photon-excited Dirac fermions manifests itself in the successive filling of the lower energy bulk states and Dirac SS and causes a broadband AB contribution extending from the initial energy of two-photon excitation (2 × 1.7 eV + $E_F$) towards the Dirac SS2 [Fig. 1(e)]. The carrier relaxation rate corresponds to the LO-phonon cascade emission (the Frohlich relaxation mechanism) [32], which manifests itself for ~1.0 ps, thus significantly exceeding the temporal chirp of the supercontinuum probing pulse [Fig. 1(f) and (g)]. This observation implies that the real sub-picosecond relaxation dynamics of two-photon-excited Dirac fermions cannot be confused with the pulse chirping effect under any experimental configuration [34].

In the real-space representation, this type of relaxation means that there must be at least two vertical transfers of carriers from the Dirac SS to the bulk states and then back to the Dirac SS. Compared to conventional semiconductors, this behavior in TIs is of decisive importance, since it leads to a gain in the mass of massless Dirac fermions due to time-reversal symmetry breaking when carriers move from the surface into the bulk. Overall, the discussed dynamics of relaxation can be considered within the framework of the Burstein-Moss effect [35-37] but being applied to all allowed electronic states [38]. Consequently, appropriate probing at a certain photon energy will make it possible to track the relaxation process through the filling of the phase-space and the corresponding Pauli blocking in the Dirac SS (the momentum space component perpendicular to the film plane is fixed) and in the bulk states (all components of the coordinate and momentum space are available). Owing to the AB nature of this relaxation dynamics, the negative transient signal is present only if the photoexcited carriers occupy the corresponding states, otherwise the amplitude of the CB-AB response would be zero. The population of higher energy states upon two-photon pumping is also evidenced by the shape of the CB-AB response. For the 3D TI $Bi_2Se_3$, the CB-AB contribution has a maximum at ~2.1 eV, thus closely corresponding to the imaginary part of the dielectric function in the visible spectral region [48]. Since the latter is known to be proportional to the CB density of states, this correlation additionally confirms the filling of the phase-space in the bulk states. The maximum of the CB-AB response is significantly shifted towards higher energies (~2.7 eV) for the 2D TI $Bi_2Se_3$. This behavior clearly demonstrates the filling of the phase-space in the Dirac surface states and the effect of their gapping [33]. This detailed consideration apparently dispels all doubts about the two-photon pumping of massless Dirac fermions.

However, two-photon pumping occurs simultaneously with the one-photon pumping of bulk bound valence electrons both from the top of the VB and from the VB spin–orbit split components [32,46]. It is expected that interband transitions between the VB spin-orbit split components and the CB states will generate less energetic carriers, which rapidly relax to the edge states and, therefore, have little effect on the overall relaxation dynamics. Alternatively, the electrons photoexcited from the VB top are more energetic and provide a longer relaxation time, thus further contributing to the CB-AB response. The corresponding holes one-photon-excited by all mentioned interband transitions rapidly relax and block out the probing optical transitions from the VB edge and the lower Dirac cone of the Dirac SS1 [Fig. 1(e)]. The resulting narrow VB-AB contribution characterizes the slow (nanosecond/microsecond range) relaxation of one-photon-excited holes in the bulk of the film [32-34]. It is worth noticing here that the effect of the pumping scattered light was completely canceled by the cross-linear-polarized configuration used in our experiments and therefore cannot be confused with the observed long-lived VB-AB contribution.

Although a decrease in pumping power has little effect on the shape of TA spectra for the 3D TI $Bi_2Se_3$, the CB-AB contribution for the 2D TI $Bi_2Se_3$ significantly increases, while the FCA contribution weakens and its higher energy edge shifts to lower energies with decreasing pumping power. The latter trends indicate that both contributions are caused by the same population of photoexcited carriers, which is redistributed between the Dirac SS and bulk states. Consequently, the relaxation dynamics in the 2D TI $Bi_2Se_3$ tends to include the bulk states to a greater extent with decreasing pumping power. On the contrary, with an increase in pumping power, relaxation through the Dirac SS manifests itself more significantly. This type of carrier redistribution is typical for ultrathin $Bi_2Se_3$ films and was also observed under one-photon UV pumping with 3.65 eV photons [33,34].

The discussed redistribution of two-photon-excited Dirac fermions between the Dirac SS and bulk states upon relaxation also manifests itself through the relaxation rates [Fig. 1(f) and (g)]. In particular, the dynamics of relaxation exhibits a two-stage behavior, which is characterized by the corresponding rates of ~1.7 and ~3.5 eV ps$^{-1}$. The difference in rates is associated with quasielastic and inelastic electron-phonon interactions in the Dirac SS and bulk states, respectively [32-34]. Furthermore, the partial contribution of the rate characterizing the relaxation of Dirac fermions in the Dirac SS increases significantly with decreasing film thickness.

To summarize this section, the TA spectra of the 3D and 2D TI $Bi_2Se_3$ films measured with low-power IR pumping show characteristics that are similar to those previously reported for more powerful pumping. The high efficiency of two-photon absorption observed even at low-power IR pumping with absorbed fluences in the μJ cm$^{-2}$ range is associated with the giant nonlinearity of Dirac fermions in Dirac systems. This observation suggests that the two-photon pumping of Dirac fermions in the TI $Bi_2Se_3$ should be taken into account in addition to the usual one-photon pumping in all pump-probe experiments performed even at low pumping power.

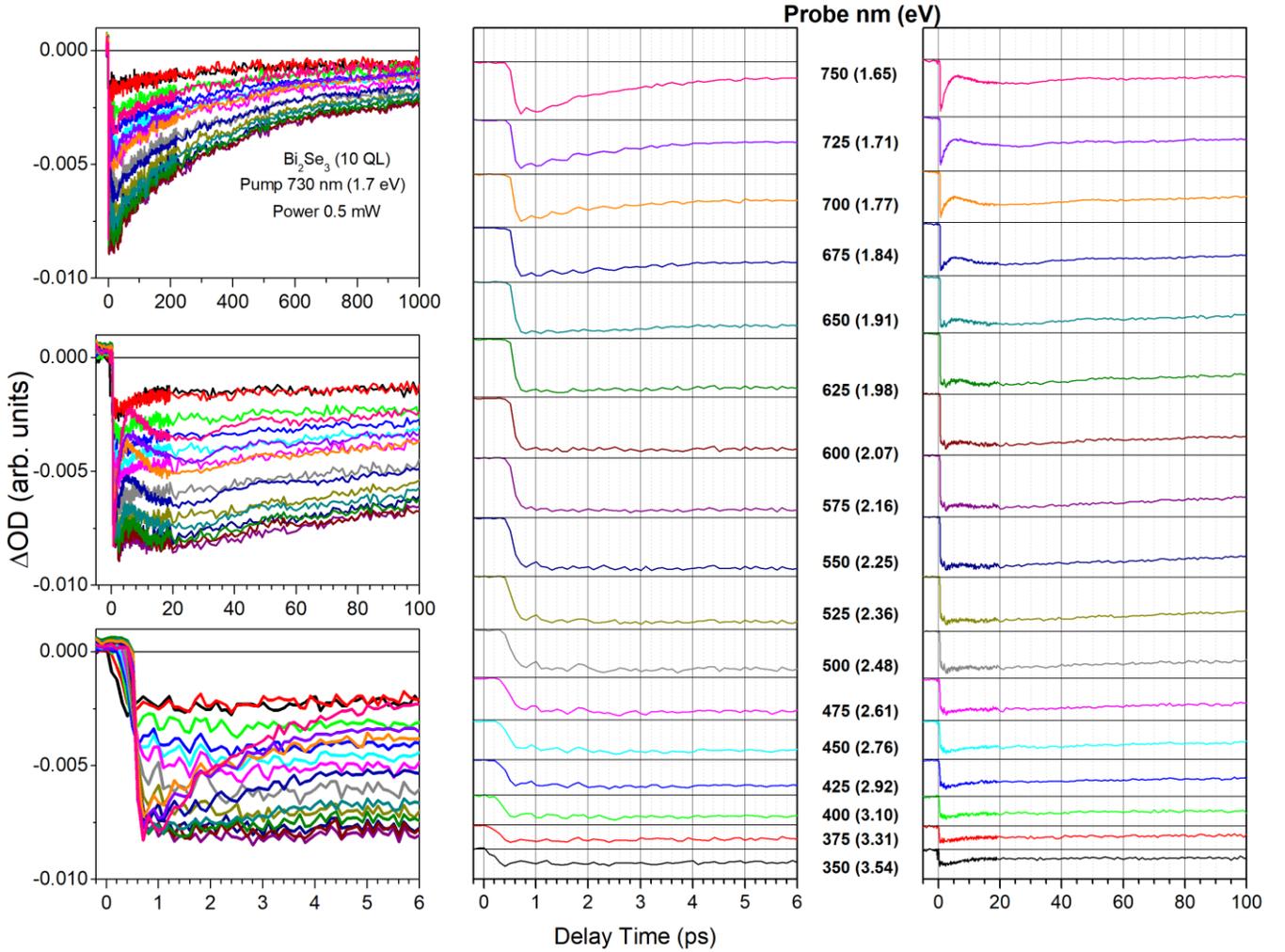

**Figure 2.** Pump-probe traces of the 3D TI $Bi_2Se_3$ (10 QL thick film) measured at different probing wavelengths (photon energies), as indicated by the corresponding colors, using the ~730 nm pumping (~1.7 eV photon energy) of ~0.5 mW average power (~4 GW cm$^{-2}$ pulse peak intensity, ~0.21 mJ cm$^{-2}$ absorbed fluence). All plots show the same set of traces plotted together or separately using different delay-time scales. The color legends are the same for all plots and are shown between the second and third columns.

*3.2. The Pump-probe traces of the 10 and 2 QL thick $Bi_2Se_3$ films.*

The corresponding CB-AB pump-probe traces represent the relaxation dynamics of carriers photoexcited by an IR pumping pulse and probed at a certain UV-Vis wavelength (photon energy). Figures 2, 3 and 4, 5 show the pump-probe traces of the 3D and 2D TI $Bi_2Se_3$ measured using the low-power and high-power pumping regimes, respectively. The traces were plotted as a function of probing photon energy and presented in different plots either together or separately using different delay-time scales for better observation.

For the 3D TI $Bi_2Se_3$, all pump-probe traces are negative due to the AB nature of the transient response [Figs. 2, 3 and 6(a) - (c)]. In addition, the pump-probe traces contain the fast-decay and long-decay stages. However, the fast-decay stage shows different decay-time constants of ~5.0 and ~1.2 ps for probing photon energies ranging from ~3.6 to ~2.5 eV and from ~2.0 to ~1.65 eV, respectively. This behavior unambiguously indicates the presence of two different relaxation dynamics. Moreover, the fast-decay relaxation dynamics of the 3.6 - 2.5 eV spectral region smoothly transforms into the long-decay dynamics with a decay-time constant of ~600 ps. In contrast, the fast-decay relaxation dynamics arising in the 2.0 - 1.65 eV spectral region is most likely transformed first into the longer dynamics including a single-cycle oscillation, which is usually associated with the excitation of coherent acoustic phonons [13,15,18], and then into the long-decay dynamics [Fig. 2, 3 and 6 (c)]. However, the fast-decay stage and the single-cycle oscillatory behavior in the latter case are not necessarily unique since both disappear when probing at photon energies exceeding ~2.0 eV. Consequently, they refer rather to the spatial redistribution of Dirac fermions upon relaxation between the Dirac SS and bulk states than to coherent acoustic phonons [32]. This conclusion is also confirmed by the fact that the fast-decay stage is completely absent when probing at photon energies ranging from ~2.0 to ~2.5 eV. The latter behavior implies that the long-decay stage completely determines the dynamics of electron relaxation in this probing region. In addition, the long-decay stage remains nearly unchanged within the entire probing region [Fig. 6(c)]. Thus, totally three relaxation dynamics can be distinguished (two fast-decay stages and a long-decay stage), which are clearly manifested in different probing regions. The observed non-monotonic variations of pump-probe traces with probing photon energy confirms a spatial redistribution of electrons upon relaxation. This spatiotemporal carrier dynamics is usually associated with the vertical electron transport [19,32].

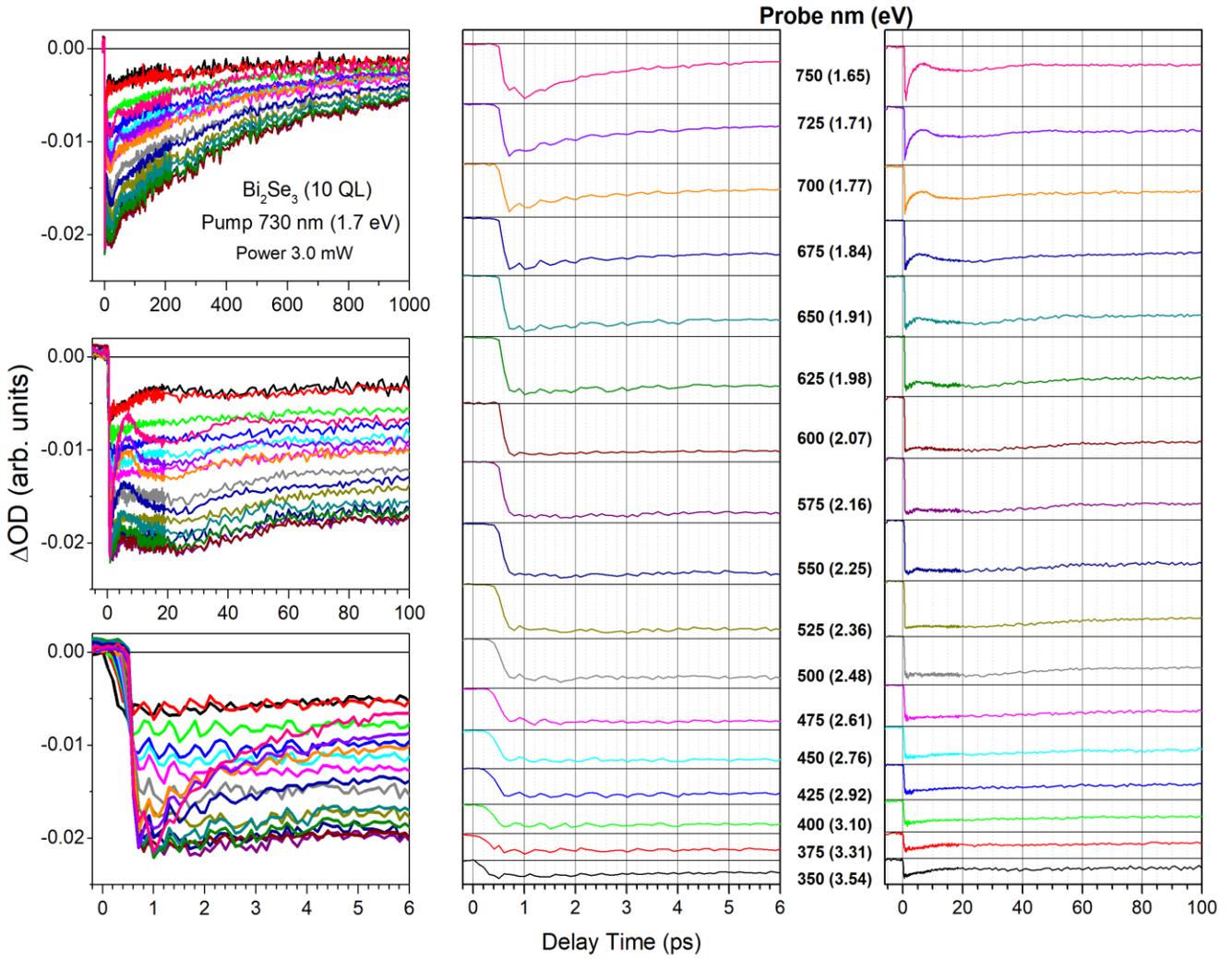

**Figure 3.** Pump-probe traces of the 3D TI $Bi_2Se_3$ (10 QL thick film) measured at different probing wavelengths (photon energies), as indicated by the corresponding colors, using the ∼730 nm pumping (∼1.7 eV photon energy) of ∼3.0 mW average power (∼27 GW cm$^{-2}$ pulse peak intensity, ∼1.0 mJ cm$^{-2}$ absorbed fluence). All plots show the same set of traces plotted together or separately using different delay-time scales. The color legends are the same for all plots and are shown between the second and third columns.

Additionally, the rise-time constant of pump-probe traces gradually shortens when probing at lower photon energies from its initial value of ~0.25 ps at ~3.6 eV to ~0.1 ps at ~2.1 eV, being stabilized afterward [Figs. 2, 3 and 6(g)]. The latter dynamics weakly depend on pumping power and is accompanied by a temporal shift in the onset of pump-probe traces, thereby demonstrating an opposite to the rise-time constant tendency with decreasing probing photon energy [Fig. 6(g)]. Here we especially emphasize that this type of correlation can be observed only if IR pumping is applied. Alternatively, for one-photon UV pumping with 3.65 eV photons, this temporal shift in the onset was also observed with decreasing probing photon energy, but the rise-time constant remained fixed [33,34]. This behavior additionally confirms the existence of two fast-decay relaxation dynamics in the IR pumped 3D TI $Bi_2Se_3$, one of which replaces the other as the probing photon energy decreases.

We associate these two fast-decay relaxation dynamics with the vertical electron transport that balances the density of photoexcited carriers between the Dirac SS and bulk states upon relaxation. On the contrary, the base amplitude of pump-probe traces (the long-decay stage) characterizes the overall relaxation tendency. To be more specific, we first consider two mechanisms that could potentially contribute to the CB-AB response. Both mechanisms are associated with Burstein-Moss effect and take into consideration phase-space filling with photoexcited carriers [35-38]. As discussed in the preceding section, phase-space filling in the Dirac SS causes the CB-AB response to appear through Pauli blocking, which occurs for massless Dirac fermions confined at the surface. Alternatively, the filling of the phase-space in the bulk states is a more general phenomenon, as it additionally involves vertical transport of carriers through the bulk states, however, eventually also leading to Pauli blocking. These mechanisms remained undistinguishable in pump-probe reflectivity experiments on the TI Bi$_2$Se$_3$, in which, due to the Kramers-Kronig relations, one can study the ultrafast carrier population dynamics using the pumping and probing beams of the same wavelength [13-19]. In contrast, these mechanisms can be observed directly in TA spectroscopy.

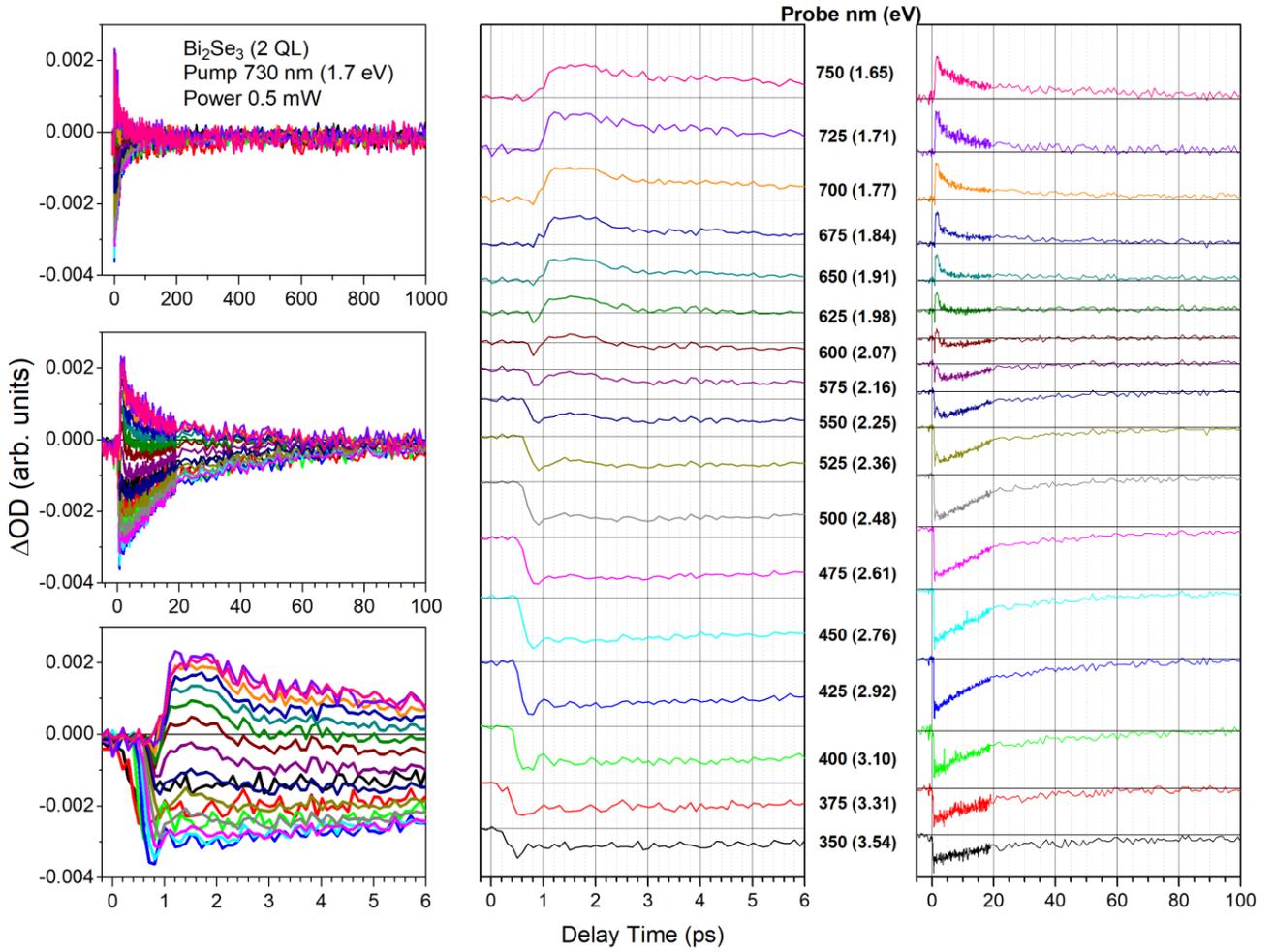

**Figure 4.** Pump-probe traces of the 2D TI Bi$_2$Se$_3$ (2 QL thick film) measured at different probing wavelengths (photon energies), as indicated by the corresponding colors, using the ∼730 nm pumping (∼1.7 eV photon energy) of ∼0.5 mW average power (∼4 GW cm$^{-2}$ pulse peak intensity, ∼16 µJ cm$^{-2}$ absorbed fluence). All plots show the same set of traces plotted together or separately using different delay-time scales. The color legends are the same for all plots and are shown between the second and third columns.

Specifically, the filling of the phase-space in the bulk states and the corresponding change in the complex refractive index is governed by electrons that were one-photon-excited from the VB top (BS1) into the CB states (BS2) and by two-photon-excited Dirac fermions relaxing to the same CB states (BS2) [Fig. 1(e)]. Further relaxation of one-photon-excited electrons to the Dirac SS2 occurs through the Frohlich relaxation mechanism that is accompanied by the vertical bulk-to-surface transport [32]. Since the energy difference between BS2 and Dirac SS2 is very small (∼0.1 eV), the relaxation of electrons occurs very quickly (∼50 fs), nevertheless leading to mass loss when they enter the Dirac SS2. The final relaxation stage of electrons residing in BS2 and massless Dirac fermions residing in the Dirac SS2 is their recombination with holes residing in the VB edge and the Dirac SS1 [17]. The latter process is responsible for the long-decay stage with a decay-time constant of ~600 ps. Here, we note that this recombination mechanism is a specific feature of ultrathin $Bi_2Se_3$ films in comparison with the bulk crystals, for which recombination in the Dirac SS1 dominates [17].

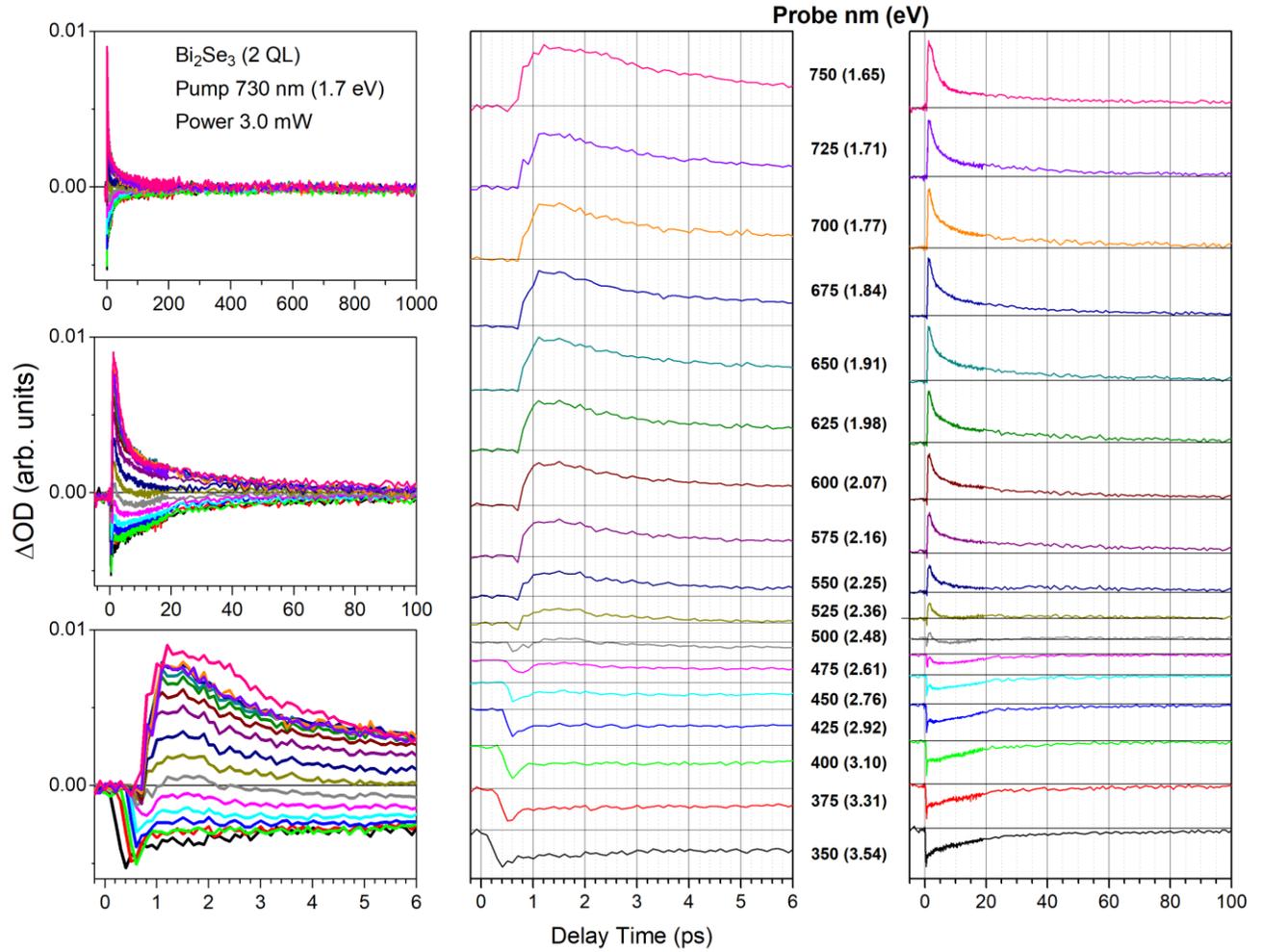

**Figure 5.** Pump-probe traces of the 2D TI $Bi_2Se_3$ (2 QL thick film) measured at different probing wavelengths (photon energies), as indicated by the corresponding colors, using the ∼730 nm pumping (∼1.7 eV photon energy) of ∼3.0 mW average power (∼27 GW cm$^{-2}$ pulse peak intensity, ∼27 μJ cm$^{-2}$ absorbed fluence). All plots show the same set of traces plotted together or separately using different delay-time scales. The color legends are the same for all plots and are shown between the second and third columns.

On the contrary, the initial effect of two-photon-excited Dirac fermions on the complex refractive index modulation in the bulk is negligible. This behavior is confirmed by the fact that the amplitude of CB-AB pump-probe traces in the probing photon energy range near two-photon pumping is minimal. Consequently, the relaxation of two-photon-excited Dirac fermions initially occurs through the Dirac SS. The corresponding rise-time constant of ~0.25 ps is a measure of Dirac fermion thermalization in the Dirac SS4 [Fig. 1(e)]. However, due to the 2D limit, the efficiency of the phase-space filling process in this case is lower than in the bulk states, as discussed in the preceding section for the relaxation rates. For the same reason, massless Dirac fermions tend to occupy the bulk states during relaxation more and more due to the higher density of states and a much higher relaxation rate for the bulk states as compared to the Dirac SS [32]. Since the Dirac surface states spatially extend from the surface into the bulk to a critical distance of ~ 3 nm (half the critical film thickness at which a gap in Dirac SS opens) [39], the surface-to-bulk vertical electron transport increases the filling of the phase-space in the bulk states that make the main contribution to the CB-AB response. Furthermore, the onset of pump-probe traces shows a temporal shift [Fig. 6(g)], since it takes some time for the relaxing Dirac fermions to reach the lower energy states, which are directly probed at certain photon energies. This assignment is also confirmed by the shape of the CB-AB response in the TA spectra measured for ~0.6 - 0.7 ps, which is still very far from the shape associated with the bulk-related density of states [48], nevertheless, clearly indicates the effect of phase-space filling in the Dirac SS4 and Dirac SS3 [Fig. 1(c)-(e)].

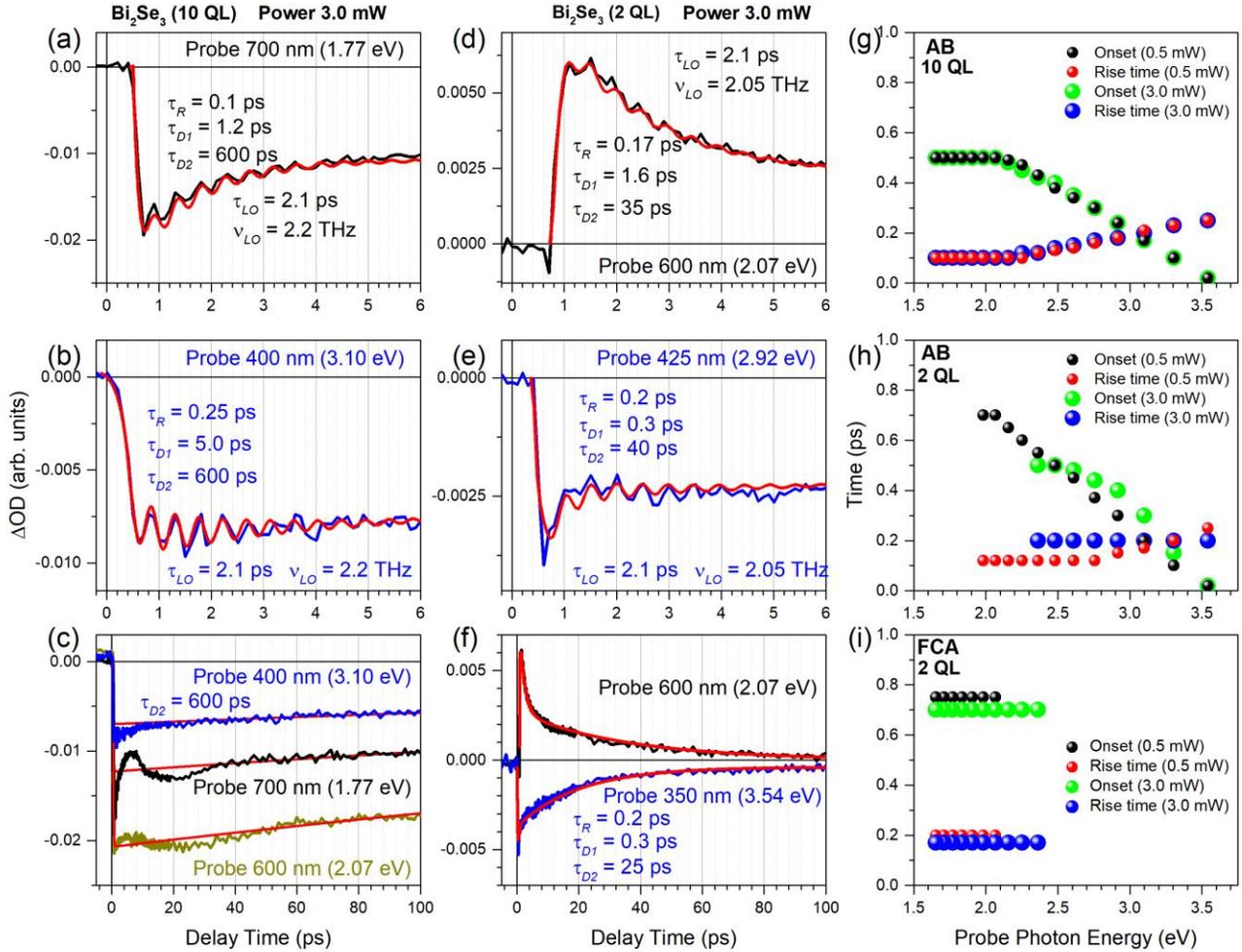

**Figure 6.** (a), (b), (c) Pump-probe traces of the 3D TI Bi$_2$Se$_3$ (10 QL thick film) measured at different probing wavelengths (photon energies), as indicated by the corresponding colors, using the ~730 nm pumping (~1.7 eV photon energy) of ~3.0 mW average power (~27 GW cm$^{-2}$ pulse peak intensity, ~1.0 mJ cm$^{-2}$ absorbed fluence). (d), (e), (f) Pump-probe traces of the 2D TI Bi$_2$Se$_3$ (2 QL thick film) measured at different probing wavelengths (photon energies), as indicated by the corresponding colors, using the ~730 nm pumping (~1.7 eV photon energy) of ~3.0 mW average power (~27 GW cm$^{-2}$ pulse peak intensity, ~27 μJ cm$^{-2}$ absorbed fluence). The corresponding rise-time and decay-time constants, as well as the results of the fast Fourier transformation of coherent LO-phonon oscillations, are listed for both columns. (g) and (h), (i) The probing photon energy dependences of the onset and rise-time constant of the CB-AB and FCA pump-probe traces measured using the low-power (0.5 mW) and high-power (3.0 mW) pumping regime for the 3D and 2D TI Bi$_2$Se$_3$, respectively.

The relaxation dynamics associated with the Dirac SS4 and Dirac SS3 and subsequent surface-to-bulk vertical electron transport manifests itself in the CB-AB pump-probe traces as the fast-decay stage in the probing range of 3.6 - 2.5 eV. The resulting CB-AB pump-probe traces are characterized by a smallest initial amplitude, which, however, steady increases when probing at lower photon energies (Fig. 2 and 3). The decay-time constant of this fast-decay stage (~5.0 ps) [Fig. 6(b)] characterizes the relaxation of two-photon-excited Dirac fermions, which includes their surface-to-bulk vertical transport.

Thus, the surface-to-bulk vertical electron transport leads to phase-space filling in the bulk states, where Dirac fermions gain masses upon reaching the topologically trivial states in which the time-reversal symmetry is broken [49]. Note that the latter process is not associated with the opening of a gap in the Dirac SS [49,50], but is due to the transfer of Dirac fermions to a distance exceeding the critical distance from the surface. For this reason, the surface-to-bulk vertical transport of Dirac fermions strongly depends on the film thickness and the density of photoexcited Dirac fermions and, therefore, can be observed mainly for the 3D TI Bi$_2$Se$_3$ (10 nm thick film). The corresponding rise-time constant of the CB-AB pump-probe traces is gradually reduced to ~0.1 ps for lower energy probing photons [Fig. 6(g)] and is determined rather by the relaxation rate of electrons than by their actual electron-electron thermalization. The discussed relaxation of nonequilibrium Dirac fermions depicts hence their transient accumulation in certain bulk and surface states situated at energies below the two-photon pumping energy of Dirac fermions (BS4, BS3, SS3, BS2 and SS2) [Fig. 1(e)]. This behavior of pump-probe traces reflects the same dynamics that is manifested in the TA spectra [Fig. 1(e)].

Once relaxing Dirac fermions transiently accumulate in BS2, the amplitude of pump-probe traces reaches its maximum and shows exclusively a long-decay behavior for the probing range of 2.0 – 2.5 eV. We associate this log-decay stage with recombination of electrons residing in BS2 and holes residing in the VB edge (BS1) [Fig. 1(e)]. However, the electron in BS2 also transfer to the Dirac SS2 [Fig. 1(e)]. This bulk-to-surface vertical electron transport manifest itself as the fast-decay stage in the probing range of ~2.0 to ~1.65 eV [Figs. 2, 3, 6(a)]. The effect of this fast-decay relaxation stage also manifests itself in a decrease in the amplitude of TA spectra at a photon energy somewhat higher than the Dirac SS2 [Fig. 1(e)]. The shorter decay-time constant of this fast-decay relaxation dynamics (~1.2 ps), as compared to that in the probing range of 3.6 - 2.5 eV (~5.0 ps), indicates a more efficient vertical electron transport to occur. This behavior also includes a single-cycle oscillation, which result from the spatial redistribution of electrons toward the Dirac SS2, followed by a recoil effect that partially move electrons back to the bulk states [32]. The recombination of carriers in the bulk states occur hence simultaneously with recombination between Dirac fermions residing in the Dirac SS2 and holes residing in Dirac SS1, as discussed above for one-photon excited carriers in the bulk. Consequently, the relaxation dynamics of two-photon-excited Dirac fermions before recombination is their transient accumulation in BS2 and the Dirac SS2, similarly to how it happens with one-photon-excited electrons in the bulk states. The transient accumulation of Dirac fermions in the Dirac SS2 has also been confirmed by higher nonequilibrium mobility of carriers found for these states, as compared to the equilibrium conductivity in the Dirac SS1 [25].

A remarkable feature of the 3D TI $Bi_2Se_3$ is the appearance of coherent LO-phonon oscillations in pump-probe traces. These oscillations are manifested exclusively at the probing photon energies, where the fast-decay stages appear (Figs. 2 and 3). On the contrary, for the probing photon energy range of 2.0 – 2.5 eV, in which the fast-decay stage and the corresponding vertical electron transport are absent, the coherent LO-phonon oscillations do not manifest themselves. Furthermore, the coherent LO-phonon oscillations reveal the same frequency of ~2.2 THz and the damping time constant of ~2.1 ps for both of above-mentioned probing regions [Fig. 6(a) and (b)], which are typical for the bulk states [7,13-16,32]. The correlation between the appearance of the fast-decay stages and the coherent LO-phonon oscillations indicates that the cooling of hot carriers due to Fröhlich interaction (the successive launching of LO-phonons) is accompanied by the LO-phonon-assisted vertical electron transport [32]. As pumping power increases, the vertical electron transport is enhanced, and the coherent LO-phonon oscillations become more pronounced. We consider the vertical transport of carriers during their relaxation as a mechanism that spatially synchronizes the electron-phonon scattering events in individual QLs, thus introducing coherence into the system. Accordingly, the coherent LO-phonon oscillations appeared in the CB-AB pump-probe traces in the probing photon energy ranges of 3.6 - 2.5 eV and 2.0 - 1.65 eV are associated with the LO-phonon-assisted surface-to-bulk and bulk-to-surface vertical electron transport, respectively.

For the 2D TI $Bi_2Se_3$, since a film thickness of ~2 nm is associated with the topologically trivial insulator phase [39,49], the observed relaxation dynamics is mainly governed by the two-photon-excited massive Dirac fermions. The spatial confinement of massive Dirac fermions also causes their high density and, consequently, the inverse bremsstrahlung type FCA in the Dirac SS2 [32]. As discussed above, this behavior leads to the appearance of additional positive contributions in the TA spectra in comparison with the 3D TI $Bi_2Se_3$ [Fig. 1(a) and (b)]. The corresponding CB-AB and FCA pump-probe traces (Figs. 4 and 5) show that the characteristic energy of probing photons, at which pump-probe traces change sign, shifts towards higher energies with increasing pumping power. This behavior is another manifestation of the same dynamics that was discussed above for TA spectra [Fig. 1(a) and (b)]. We also note that the CB-AB and FCA pump-probe traces for the 2D TI $Bi_2Se_3$ reveal much shorter decay-time constants of the long-decay stage as compared to the 3D TI $Bi_2Se_3$ [Fig. 6(a) - (f)]. This observation confirms the general tendency, according to which the rate of electron-hole recombination increases with decreasing film thickness [17].

For the low-power pumping regime applied to the 2D TI $Bi_2Se_3$, the temporal shift in the onset of the CB-AB pump-probe traces and a decrease in their rise-time constants with decreasing probing photon energy are very similar to those observed for the 3D TI $Bi_2Se_3$ [Fig. 6(g) and (h)]. However, as pumping power increases, the rise-time constant of pump-probe traces becomes independent of probing photon energy, thereby demonstrating a similar trend observed for one-photon UV pumping [33,34]. This behavior suggests that for the 2D TI $Bi_2Se_3$, the filling of the phase-space in the bulk states becomes negligible with increasing pumping power. Consequently, the relaxation dynamics of massive Dirac fermions in this case mainly appeared through the filling of the phase-space in the Dirac SS.

One of the most interesting dynamics, which manifest itself in the pump-probe traces of the 2D TI $Bi_2Se_3$, is associated with the overlap of positive and negative contributions. Since the positive contribution characterizes FCA in the Dirac

SS2 and, therefore, arises exclusively when the two-photon-excited massive Dirac fermions transiently accumulate in these states, the onset of the FCA pump-probe traces shows a significant temporal shift. This behavior appears as a negative narrow feature in the CB-AB pump-probe traces on a subpicosecond time scale [Fig. 6(e)], which is caused by the overlap of the negative and positive contributions. It is important to note that as soon as the FCA pump-probe traces appear at a certain photon energy, there is no noticeable temporal shift in their onset and no change in the rise-time constant with further decrease in probing photon energy [Fig. 6 (i)]. This relaxation dynamics clearly indicates that the accumulation of massive Dirac fermions in the gapped Dirac SS2 is the final relaxation stage before recombination.

The temporal overlap of the CB-AB and FCA pump-probe traces also influences the appearance of coherent LO-phonon oscillations. The vertical transport of massive Dirac fermions in the 2D TI $Bi_2Se_3$ is significantly weakened. As a result, coherent LO-phonon oscillations appear exclusively in the FCA pump-probe traces [Figs. 4, 5, and 6(d), (e)]. However, because of the overlap of the positive and negative contributions, the CB-AB pump-probe traces show coherent LO-phonon oscillations that replicate those associated with the FCA response. This tendency is confirmed by the lower frequency of coherent LO-phonon oscillations in the Dirac SS (~2.05 THz) [7,32], and the constant phase of oscillations for all traces observed at different probing photon energies. This behavior is in complete contrast with the phase-dependent coherent surface-to-bulk vibrational coupling observed for a 2 nm thick $Bi_2Se_3$ film under one-photon UV pumping with 3.65 eV photons [34]. This discrepancy illustrates the loss of coherence of LO-phonon oscillations in the bulk under simultaneous one-photon and two-photon pumping. The launching of coherent LO-phonon oscillations in the Dirac SS2 is likely governed by the in-phase successive quasielastic scattering of massive Dirac fermions from the uppermost atomic layer of the films. The in-phase launching in this case means that the sources of the Dirac fermion scattering are not spatially separated, as that occurs in the bulk states.

## 4. Conclusions

In summary, using UV-Vis TA spectroscopy, we have shown that in pump-probe experiments on the 2D and 3D TI $Bi_2Se_3$ exploiting IR ultrashort (100 fs) pumping pulses with a wavelength of 730 nm (1.7 eV photon energy)], even at low-power pumping with absorbed fluences in the µJ cm$^{-2}$ range, TA spectra extend across a part of the UV and the entire visible region. This observation unambiguously proves the two-photon pumping regime. We attributed the high efficiency of two-photon pumping to the giant nonlinearity of Dirac fermions in the Dirac SS. Alternatively, one-photon pumping is associated with the excitation of bulk bound valence electrons into the conduction band. Two mechanisms of absorption bleaching have been identified, which we associate with the filling of the phase-space in the Dirac SS and bulk states and the corresponding Pauli blocking. We suggested that the coherent launching of LO-phonon oscillations in the bulk states and Dirac SS is due to the vertical electron transport and the in-phase sequential quasielastic scattering of Dirac fermions from the uppermost atomic layer, respectively. Finally, we concluded that the two-photon IR pumped UV-Vis TA spectroscopy of Dirac fermions is a powerful and selective tool for studying ultrafast relaxation dynamics in 2D Dirac systems.

**Author Contributions:** Y.D.G. modified and tested the transient absorption spectrometer, built the experimental setup, performed optical measurements, and treated the optical experimental data. The optical measurements were performed in the laboratory hosted by T.H. All authors contributed to discussions. Y.D.G. analyzed the data and wrote this paper. X.W.S. guided the research and supervised the project.

**Acknowledgements:** This work was supported by the National Key Research and Development Program of China administrated by the Ministry of Science and Technology of China (Grant No. 2016YFB0401702), the Shenzhen Peacock Team Project (Grant No. KQTD2016030111203005), and the Shenzhen Key Laboratory for Advanced Quantum Dot Displays and Lighting (Grant No. ZDSYS201707281632549). The authors acknowledge J. Li for help with the laser system operation and S. Babakiray for growing the $Bi_2Se_3$ samples by MBE (under supervision of D. Lederman) using the West Virginia University Facilities.

**Data Availability Statement:** The data presented in this study are available on request from the corresponding author.

**Conflicts of Interest:** The authors declare no conflict of interest.